\documentclass[showpacs,amsmath,amssymb,twocolumn,prl,superscriptaddress]{revtex4-1}
\usepackage{amssymb}
\usepackage[dvips]{graphicx}
\usepackage{enumerate}
\usepackage{epsfig}
\usepackage{subfigure}
\usepackage{xcolor}
\usepackage[T1]{fontenc}
\usepackage{fullpage}
\usepackage{amsthm,amsfonts,amssymb,amscd,mathrsfs,xspace,framed}
\usepackage{amsmath}
\usepackage{color}
\usepackage{setspace}
\usepackage{url}
\usepackage{wrapfig}
\usepackage{enumitem}
\newcommand{\defeq}{\overset{\text{\tiny def}}{=}}
\begin{document}

\title{High-dimensional Frequency-Encoded Quantum Information Processing with Passive Photonics and Time-Resolving Detection}

\author{Chaohan Cui}
\affiliation{James C. Wyant College of Optical Sciences, The University of Arizona, Tucson, Arizona 85721, USA}
\author{Kaushik P. Seshadreesan}
\affiliation{James C. Wyant College of Optical Sciences, The University of Arizona, Tucson, Arizona 85721, USA}
\author{Saikat Guha}
\affiliation{James C. Wyant College of Optical Sciences, The University of Arizona, Tucson, Arizona 85721, USA}
\author{Linran Fan}
\email{lfan@optics.arizona.edu}
\affiliation{James C. Wyant College of Optical Sciences, The University of Arizona, Tucson, Arizona 85721, USA}

\begin{abstract}
In this Letter, we propose a new approach to process high-dimensional quantum information encoded in a photon frequency domain. In contrast to previous approaches based on nonlinear optical processes, no active control of photon energy is required. Arbitrary unitary transformation and projection measurement can be realized with passive photonic circuits and time-resolving detection. A systematic circuit design for a quantum frequency comb with arbitrary size has been given. The criteria to verify quantum frequency correlation has been derived. By considering the practical condition of detector's finite response time, we show that high-fidelity operation can be readily realized with current device performance. This work will pave the way towards scalable and high-fidelity quantum information processing based on high-dimensional frequency encoding.
\end{abstract}

\maketitle
A scalable approach to generate and control large-scale quantum systems is critical to realize meaningful quantum information processing~\cite{o2009photonic,gisin2007quantum,pirandola2018advances,weedbrook2012gaussian}. While increasing the the number of sub-systems is required to realize the exponential scaling in quantum information processing, the implementation of multiple-level sub-systems nevertheless boosts the efficiency for increasing the Hilbert space size of quantum systems.~\cite{zhou2003quantum,lanyon2009simplifying, babazadeh2017high,scalability}. The photon frequency degree of freedom has provided an ideal platform for realizing high-dimensional encoding of quantum states, due to its intrinsic multimode property~\cite{kues2019quantum,pfister2019continuous}. The classical optical comb can provide hundreds of frequency modes~\cite{cundiff2003colloquium,gaeta2019photonic}, and quantum correlation among more than 60 frequency modes has been demonstrated~\cite{chen2014experimental}. Furthermore, both continuous-variable and discrete-variable quantum frequency combs have been realized~\cite{pinel2012generation,chen2014experimental,kues2019quantum}. Multiple platforms have been developed, including optical parametric oscillators~\cite{pinel2012generation,chen2014experimental}, filtered parametric down-conversion~\cite{xie2015harnessing}, and four-wave mixing with nanophotonic ring cavities~\cite{reimer2016generation,kues2017chip, imany201850}.

While the generation of high-dimensional quantum states in a frequency domain has achieved great progress, the capability to manipulate these states is still limited. Coherent control of photon frequency modes requires active nonlinear optical processes to change photon energy. four-wave mixing was first utilized to realize frequency beam splitters~\cite{clemmen2016ramsey}. However, the manipulation beyond two modes has not been demonstrated as a large number of phase-locked lasers are required. Recently, a new approach based on electro-optic modulation and optical phase shaping was demonstrated to realize frequency beam splitters and tritters~\cite{lu2018electro}. However, the extension to multimodes also requires a complex modulation scheme by stacking multiple modulators and pulse shapers, and the design of such devices relies on extensive optimization processes~\cite{lukens2017frequency}. More importantly, this approach has intrinsic loss due to sidebands outside desirable frequency range~\cite{lukens2017frequency,lu2018electro}. Practically, the limited modulation frequency also introduces extra loss as high-order sidebands are normally used~\cite{kues2017chip}. The contradiction between the capabilities to generate and manipulate these high-dimensional quantum states has thus placed an urgent need for new approaches to process quantum information encoded in photon frequency degree of freedom.

In this Letter, we provide a novel approach to process high-dimensional quantum states encoded in photon frequency domain without active nonlinear optical processes. Frequency modes are converted into spatial modes with passive photonic circuits. Therefore unitary transformation of spatial modes will also be applied to frequency modes, and time-resolving detection of spatial modes performs the projection measurement of frequency modes. The derivation of our proposed approach focuses on discrete-variable frequency combs in the single-photon regime, detected with ultra-fast single-photon detectors. The influence of finite response time on measurement fidelity is estimated, showing that high fidelity is within reach of current technology. Based on this approach, we further propose a new method to verify quantum correlation in a frequency domain.

As well-defined discrete frequency modes are generally used for high-dimensional frequency-encoded single-photon quantum states, we start with a quantized optical field consisted of $N$ frequency modes
\begin{equation}
\begin{aligned}
\label{eq:sp}
    |\psi\rangle=&\sum_{k=0}^{N-1} c_k \hat{a}^\dag_{\omega_k}|0\rangle^{\otimes N}
\end{aligned}
\end{equation}  
with $c_k$ the complex amplitude and $\hat{a}^\dag_{\omega_k}$ the creation operator for the $k$th frequency mode. Such states have been generated with parametric down-conversion and four-wave mixing processes in the low pump power regime, where multi-pair processes are neglected  ~\cite{xie2015harnessing, reimer2016generation,kues2017chip, imany201850}. For such a discrete Hilbert space, we can define a complete set of projection operators $\{\textbf{M}_j=|M_j\rangle\langle M_j|\}$ with $|M_j\rangle=\sum_{k=0}^{N-1}u_{jk}\hat{a}^\dag_{\omega_k}|0\rangle^{\otimes N}$ and $u_{jk}$ the $jk$ component of unitary matrix \textbf{U}. These projection operators show the process of a unitary transformation \textbf{U} of a high-dimensional frequency-encoded quantum state followed by a projection measurement in the new eigen basis.
\begin{figure}[tb]
\centering
\includegraphics[width=8cm]{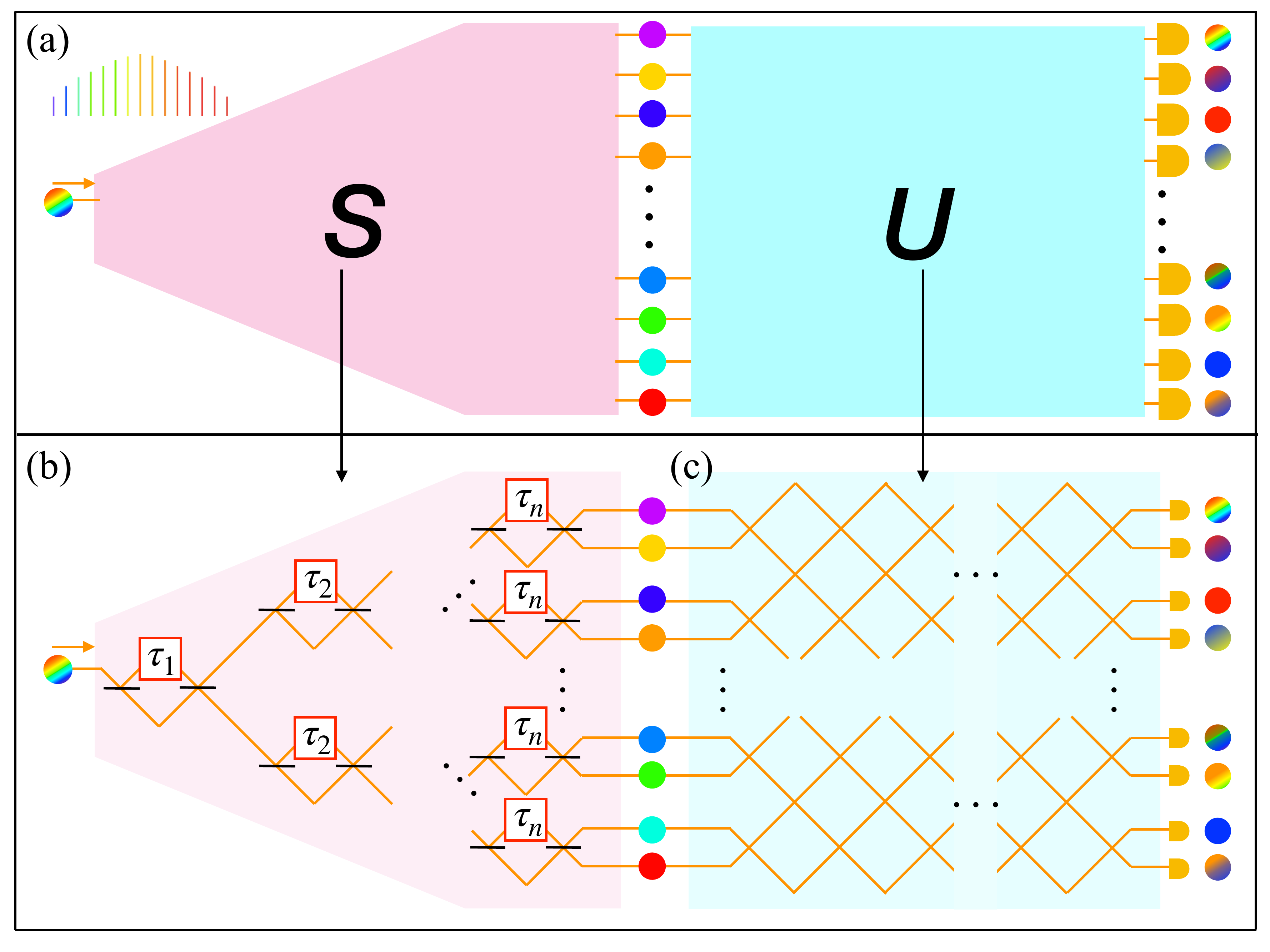}
\caption{\label{fig:SU} (a) High-dimensional frequency-encoded unitary transformation and projection measurement with passive photonic circuit and time-resolving detection. Different colors indicate different frequency modes, and the gradient color shows a superposition of multiple frequency modes. (b) Photonic circuit to realize \textbf{S} for a quantum frequency comb with evenly distributed frequency modes, $\tau_j=\pi/(2^{(j-1)}\delta\omega)$. (c) Photonic circuit to realize unitary transformation \textbf{U}.}
\end{figure}
In contrast to traditional methods utilizing nonlinear optical processes to build frequency-domain beam splitters, we show that these operations can be realized with passive photonic circuits and time-resolving detection [Fig.~\ref{fig:SU}(a)]. First, discrete frequency modes are associated with different spatial modes through operator \textbf{S}, which can practically be realized by a prism, optical grating, or Mach-Zehnder interferometer (MZI) array 
\begin{equation}
\begin{aligned}
\label{eq:S}
    &\textbf{S}|\psi\rangle=\sum_{k=0}^{N-1} c_k \hat{a}^\dag_{\omega_k,r_k}|0\rangle^{\otimes N^2}
\end{aligned}
\end{equation}
where $r_k$ is the spatial mode that carries the $\omega_k$ frequency component. For a quantum frequency comb with evenly distributed frequency modes, one possible design of the passive photonic circuit is shown in Fig.~\ref{fig:SU}(b). For the case the frequency mode number $N=2^n$ with $n$ a positive integer, the operator \textbf{S} can be realized with $2^n-1$ unbalanced MZIs~\cite{huntington2005demonstration} arranged hierarchically into $n$ levels [Fig. 1(b)]. For the $j$th level, the relative time delay of the MZIs is set to be $\tau_j=\pi/(2^{(j-1)}\delta\omega)$, with $\delta\omega$ the free-spectral range of the quantum frequency comb~\cite{CombCondition}. For the case $2^{n-1}<N<2^n$, the same circuit can be used and the redundant modes can be dropped. Then the $N$ spatial modes are sent into an $N$-port passive photonic circuit to realize unitary transformation \textbf{U} in a spatial domain~\cite{reck1994experimental,clements2016optimal,zhang2019quantum}. Since spatial modes simultaneously represent frequency modes, the interference in a spatial domain also results in mixing of associated frequency modes
\begin{equation}
\begin{aligned}
\label{eq:U}
    \textbf{U}~ \hat{a}^\dag_{\omega_k,r_k}=\sum_{j=0}^{N-1} u_{jk} \hat{a}^{(\text{out})\dag}_{\omega_k,r_j}
\end{aligned}
\end{equation}
where $r_j$ indicates the $j$th output spatial mode of the photonic circuit [Fig.~\ref{fig:SU}(c)]. Then the output spatial modes are measured with ultrafast detectors. The outcome of $j$th output port is expressed as
\begin{equation}
\begin{aligned}
\label{eq:spd}
    P_j(t;T)=&\int^{t+T/2}_{t-T/2}\frac{d\tau}{T}~\langle\psi|\textbf{S}^\dag\textbf{U}^\dag\hat{E}^{\dag}_{r_j}(\tau)\hat{E}_{r_j}(\tau)\textbf{U}\textbf{S}|\psi\rangle\\
    =&\int^{t+T/2}_{t-T/2}\frac{d\tau}{T}~\langle\psi|e^{i\hat{H}\tau/\hbar}\textbf{M}_je^{-i\hat{H}\tau/\hbar}|\psi\rangle\\
    =&\int^{t+T/2}_{t-T/2}\frac{d\tau}{T}~\langle\psi|\textbf{M}_j(\tau)|\psi\rangle\\
\end{aligned}
\end{equation}
with $T$ the detector's response time, $\hat{H}=\sum_{k=0}^{N-1}\hbar\omega_k(\hat{a}^\dag_{\omega_k}\hat{a}_{\omega_k}+\frac{1}{2})$, and $\hat{E}_{r_j}(\tau)=\sum_{k=0}^{N-1}\hat{a}^{(\text{out})}_{\omega_k,r_j}e^{-i\omega_k\tau}$, representing the equivalent photon phase change due to detection jitter. With infinitely small $T$, the measurement outcome in a spatial domain approaches the desired projection measurement in a frequency domain. 
\begin{equation}
\begin{aligned}
\lim_{T\rightarrow 0^+}P_j(t;T)=\text{Tr}(\textbf{M}_j|\psi\rangle\langle\psi|)
\end{aligned}
\end{equation}
Therefore, the complete set of projection operators in a frequency domain can be realized with passive photonic circuits and time-resolving detection. This approach is based on the Fourier correspondence between frequency and temporal correlations~\cite{pe2005temporal,o2011observations,lukens2013demonstration,bernhard2013shaping}.

Considering the practical condition of detector's finite response time, the fidelity of the projection measurement can be expressed as 
\begin{equation}
\begin{aligned}
\label{eq:Fidelity}
    &F_j=\dfrac{\left|\text{Tr}\left( \widetilde{\textbf{M}}_j(t; T)^\dag \textbf{M}_j\right)\right|^2}{\text{Tr}(\widetilde{\textbf{M}}_j(t; T)^\dag \widetilde{\textbf{M}}_j(t; T))\text{Tr}(\textbf{M}_j^\dag \textbf{M}_j)}\\
\end{aligned}
\end{equation}
with $\widetilde{\textbf{M}}_j(t; T)=\int^{t+T/2}_{t-T/2}\frac{d\tau}{T} \textbf{M}_j(\tau)$ the effective projection.
\begin{figure}[tb]
\centering
\includegraphics[width=8cm]{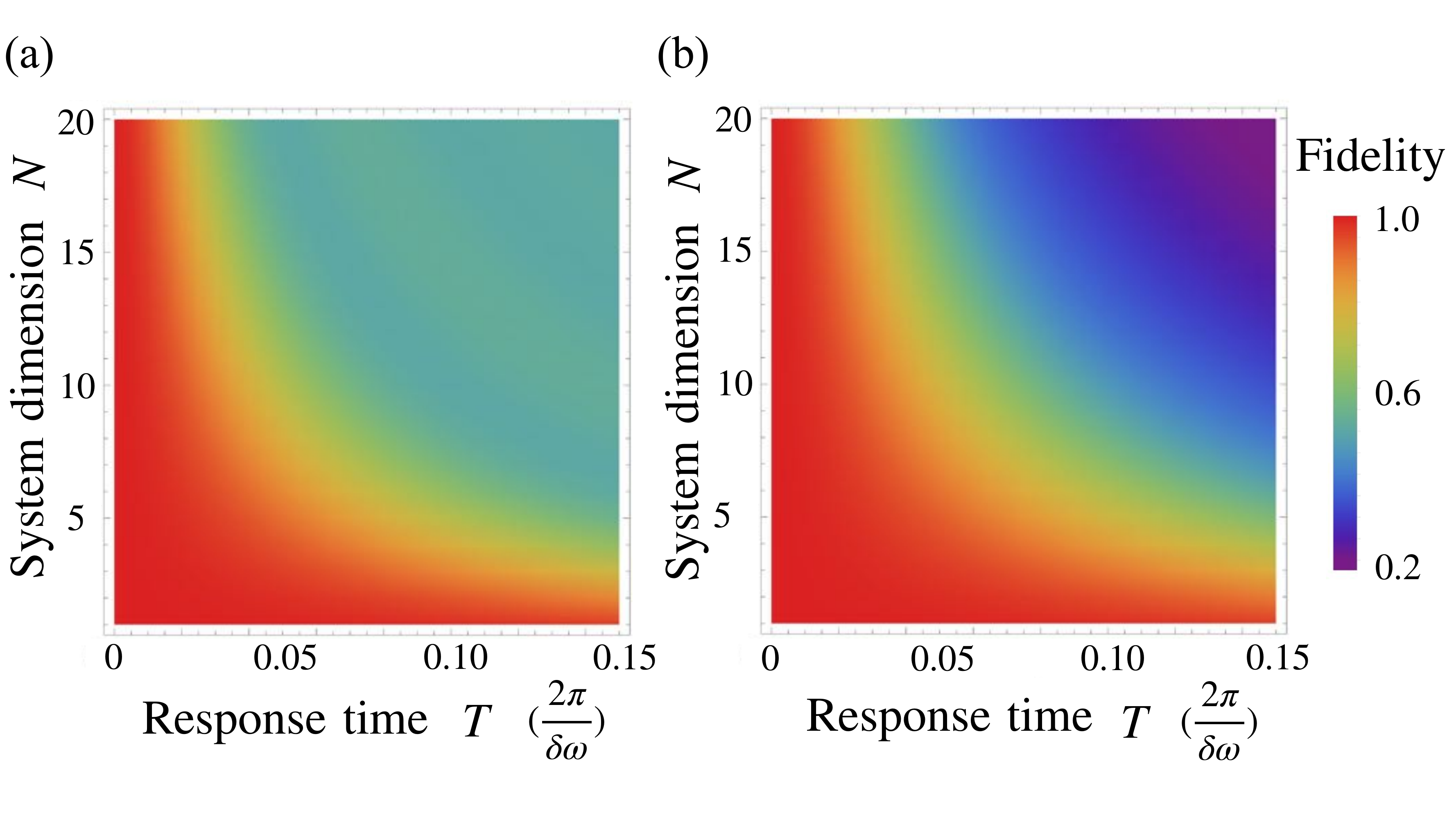}
\caption{\label{fig:Fidelity} The measurement fidelity with finite response time $T$ and dimension $N$ when the projection measurement is onto (a) the two-mode superposition with largest frequency difference $1/\sqrt{2}(\hat{a}^\dag_{\omega_0}+\hat{a}^\dag_{\omega_0+(N-1)\delta\omega})|0\rangle^{\otimes N}$ and (b) the Fourier basis $1/\sqrt{N}\sum_{k=0}^{N-1}\hat{a}^\dag_{\omega_0+k\delta\omega}|0\rangle^{\otimes N}$.}
\end{figure}
Practically, the system's total response time $T$ is dominated by phase instability in the photonic circuit or detector jitter. A large response time will inevitably degrade the fidelity of projection measurement (Fig.~\ref{fig:Fidelity}). In order to achieve high-fidelity operation, the response time needs to be much smaller than the beating period of the quantum frequency comb, $T<<2\pi/\delta\omega$. In addition, the fidelity of projection measurement with different bases have different dependencies on the response time $T$. The projection onto the basis with fast temporal evolution, such as two-mode superposition with large frequency difference [Fig.~\ref{fig:Fidelity}(a)] and the Fourier basis [Fig.~\ref{fig:Fidelity}(b)], will be more sensitive to the error induced by the finite response time. In contrast, the fidelity of the projection measurement onto the computational basis with single frequency modes will not be influenced.

With the capability to realize arbitrary projection measurement, we further develop the approach to verify bipartite  high-dimensional quantum entanglement encoded in a frequency domain with passive photonic circuits and time-resolving detection~\cite{kaszlikowski2000violations,highdimensionCHSH}. Two sets of projection bases, time-delayed Fourier basis and time-delayed inverse Fourier basis, can be defined for photon $a$ and $b$, respectively.
\vspace{2mm}
\begin{equation}
\begin{aligned}
\label{eq:basis}
    |M&_{a,j}(t)\rangle\\
    &=\frac{1}{\sqrt{N}}\sum_{k=0}^{N-1}\exp\left\{i\left(\frac{2\pi j}{N}+\delta\omega t\right)k\right\}\hat{a}^\dag_{\omega_k}|0\rangle^{\otimes N}\\
    |M&_{b,j}(t)\rangle\\
    &=\frac{1}{\sqrt{N}}\sum_{k=0}^{N-1}\exp\left\{i\left(-\frac{2\pi j}{N}+\delta\omega t\right)k\right\}\hat{b}^\dag_{\omega_{N-1-k}}|0\rangle^{\otimes N}
\end{aligned}
\end{equation}
where integer $j$ labels the index of output port. Both sets of projection measurement can be realized with the photonic circuits shown in Fig. 1. The generalized high-dimensional entanglement indicator can be written as 
\vspace{3mm}
\begin{widetext}
\begin{equation}
\begin{aligned}
\label{eq:d-Bell}
    S_N(t_a,t_b,t'_a,t'_b)\defeq\sum_{k=0}^{[N/2]-1}&(1-\frac{2k}{N-1})\{[P(t_a,t_b,k)+P(t'_a,t_b,-k-1)+P(t'_a,t'_b,k)+P(t_a,t'_b,-k)]\\
    &-[P(t_a,t_b,-k-1)+P(t'_a,t_b,k)+P(t'_a,t'_b,-k-1)+P(t_a,t'_b,k+1)]\}
\end{aligned}
\end{equation}
\begin{equation}
\begin{aligned}
    P(t_a,t_b,k)=\frac{1}{T^2}\iint_{-T/2}^{T/2} d t d t' \sum_{l=0}^{N-1}|\langle M_{a,l}(t_a+t),M_{b,(l+k)\text{mod} d}(t_b+t')|\psi\rangle|^2
\end{aligned}
\end{equation}
\end{widetext}
with $t_a$, $t'_a$ and $t_b$, $t'_b$ corresponding to different settings of the photonic circuits, thus different projections bases, for photon $a$ and $b$, respectively~\cite{highdimensionCHSH}. Classically, the maximum value of $S_N$ is 2. The quantum limit of $S_N$, which is above 2, can be achieved with the high-dimensional Bell state
\vspace{2mm}
\begin{equation}
\label{eq:state}
    |\psi\rangle=\frac{1}{\sqrt{N}}\sum_{k=0}^{N-1}\hat{a}^\dag_{\omega_k}\hat{b}^\dag_{\omega_{N-1-k}}|00\rangle^{\otimes N}.
\end{equation}
by setting the detection time $t_b-t_a=\frac{\pi}{2N\delta\omega}$ and $t'_a-t_a=t'_b-t_b=\frac{\pi}{N\delta\omega}$.
\begin{figure}[tb]
\centering
\includegraphics[width=8cm]{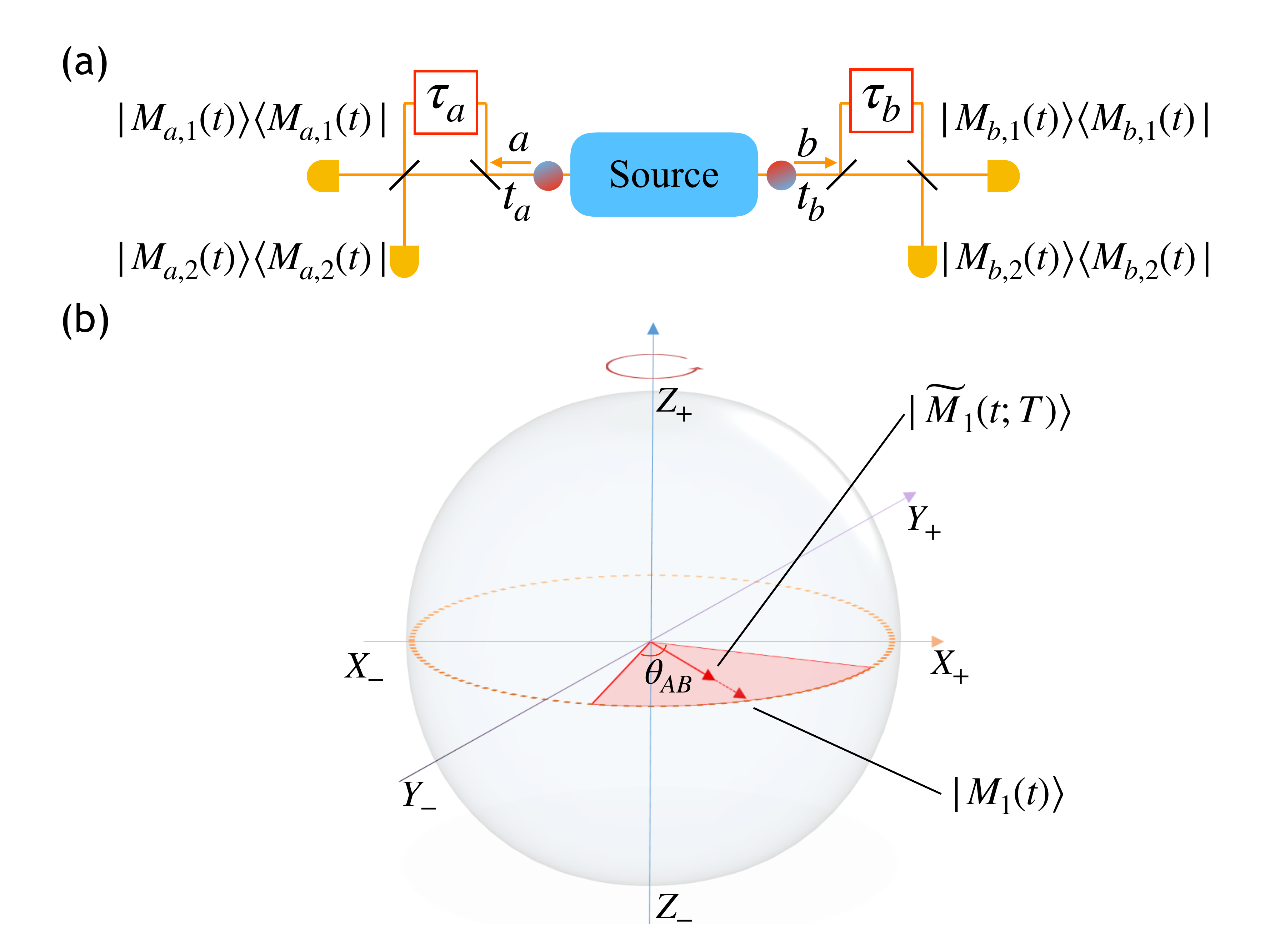}
\caption{\label{fig:Bloch} (a) Setup to verify two-frequency bipartite entanglement. (b) Bloch sphere for two-frequency systems. Red shaded area is the time averaging range with the designed and equivalent measurement shown as arrows. }
\end{figure}
While the exact value of $S_N$ is dependent on the size $N$ of high-dimensional systems~\cite{highdimensionCHSH}, we use the two-frequency case as an example to illustrate our approach and the effect of a finite response time. In two-frequency systems, the photonic circuits shown in Fig.~\ref{fig:SU} can be simplified to a single unbalanced MZI for each photon [Fig.~\ref{fig:Bloch}(a)], and the corresponding projection basis can be represented on Bloch sphere [Fig.~\ref{fig:Bloch}(b)]. With two frequency modes $\omega_1$ and $\omega_2$, the criterion for quantum entanglement can be simplified to the CHSH inequality, which can be maximally violated with the Bell state
\begin{equation}
\begin{aligned}
\label{eq:CHSHState}
    |\psi\rangle=\frac{1}{\sqrt{2}}(\hat{a}^\dag_{\omega_1}\hat{b}^\dag_{\omega_2}+\hat{a}^\dag_{\omega_2}\hat{b}^\dag_{\omega_1})|00\rangle^{\otimes N}
\end{aligned}
\end{equation}

\begin{figure}[tb]
\centering
\includegraphics[width=8cm]{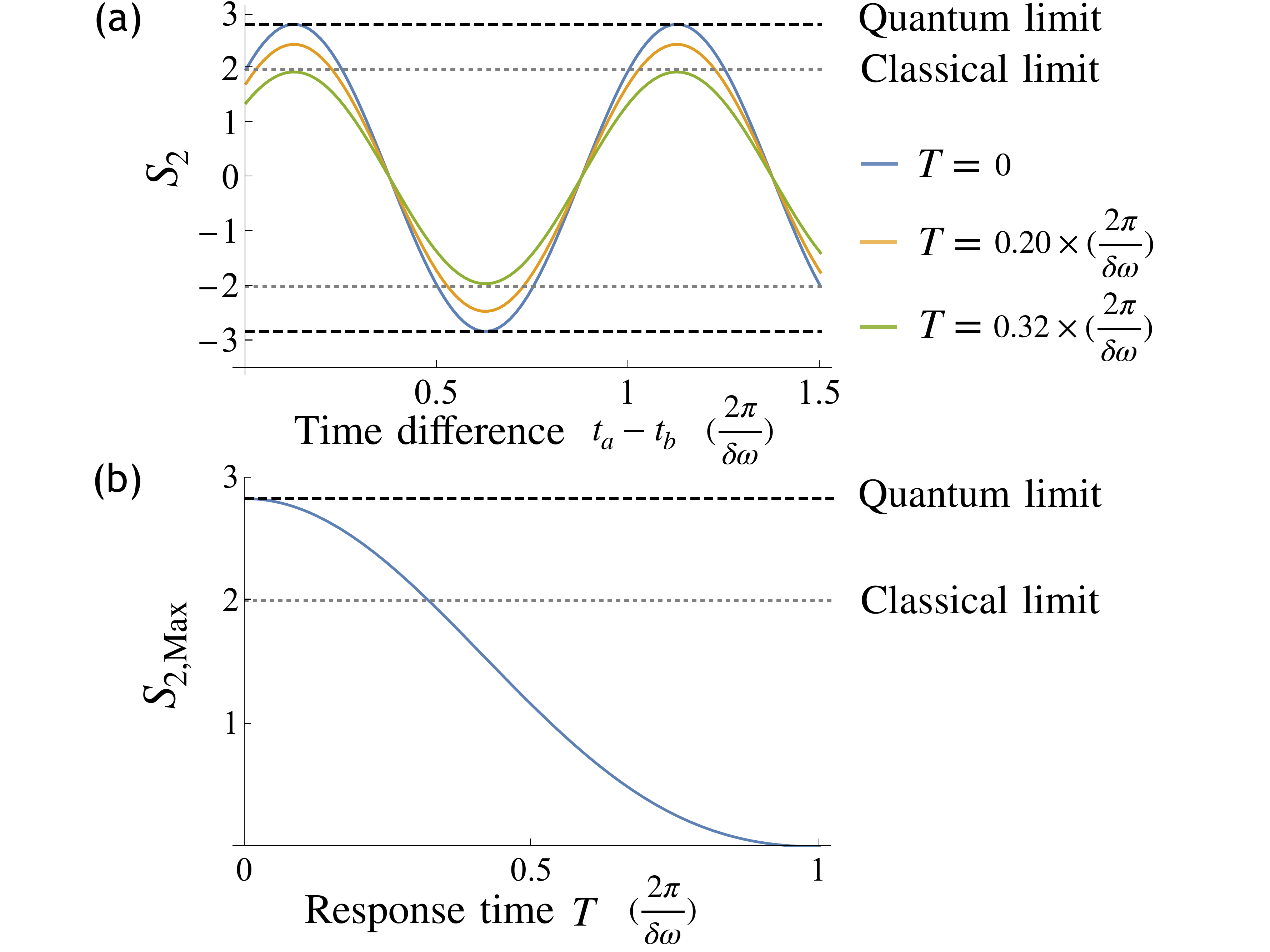}
\caption{\label{fig:CHSH} (a) CHSH quantity $S_2$ values with varying detection time difference $t_a-t_b$ and fixed $\tau_a$ and $\tau_b$. (b) Maximal measurable $S_{2,{\text{Max}}}$ with dependence on response time $T$.}
\end{figure}

By setting both relative delay $\tau_a$ and $\tau_b$ giving $\pi/2$ and -$\pi/2$ phase difference for mode $\omega_1$ and $\omega_2$ respectively~\cite{CombCondition}, the projection bases are placed on the equator of the Bloch sphere,
\begin{equation}
\begin{aligned}
\label{eq:CHSHState2}
    |M_{c,j}(t)\rangle=\frac{1}{\sqrt{2}}(\hat{c}^\dag_{\omega_1}+(-1)^j e^{i\delta\omega t_c}\hat{c}^\dag_{\omega_2})|0\rangle^{\otimes N}
\end{aligned}
\end{equation}
where subscripts $\hat{c}=\hat{a},\hat{b}$ and $j=1,2$ label the photon and measurement output, respectively. Photon clicks at output 1 and 2 correspond to $+1$ and $-1$ in the original CHSH scheme respectivley \cite{aspect1982experimental,rarity1990experimental,ou1990observation}. Then the CHSH value can be defined through the correlation coefficient $E(t_a,t_b)$ and the coincidence rate $\bar{C}_{ij}(t_a,t_b)$
\begin{widetext}
\begin{equation}
\begin{aligned}
&S_2=E(t_a,t_b)-E(t_a,t'_b)+E(t'_a,t_b)+E(t'_a,t'_b)\\
&E(t_a,t_b)=\dfrac{\bar{C}_{11}(t_a,t_b)+\bar{C}_{22}(t_a,t_b)-\bar{C}_{12}(t_a,t_b)-\bar{C}_{21}(t_a,t_b)}{\bar{C}_{11}(t_a,t_b)+\bar{C}_{22}(t_a,t_b)+\bar{C}_{12}(t_a,t_b)+\bar{C}_{21}(t_a,t_b)}\\
&\bar{C}_{ij}(t_a,t_b)\propto\frac{1}{T^2}\iint_{-T/2}^{T/2} d t d t' |\langle M_{a,i}(t_a+t),M_{b,j}(t_b+t')|\psi\rangle|^2
\end{aligned}
\end{equation} 
\end{widetext}
The maximum value of $S_2$ is achieved when photon detection time satisfies $t_b-t_a=\frac{\pi}{4\delta\omega}$, $t'_a-t_a=\frac{\pi}{2\delta\omega}$ and $t'_b-t_b=\frac{-\pi}{2\delta\omega}$. These conditions correspond to fixing the angle between projection bases for photon $a$ and $b$ to be 45 degree, and the bases are rotated by 90 degree along the Z-axis of the Bloch sphere between the two measurements. The finite response time will inevitably degrade the measurement results. In this particular setup, the effect of the finite response time corresponds to averaging along an arc on the the Bloch sphere equator [Fig.~\ref{fig:Bloch}(b)]. And the angle of the arc is equal to $T\cdot\delta\omega$. The calculated result of $S_2$ with fixed optimum $t'_a-t_a$, $t'_b-t_b$ and varying $t_b-t_a$ values is shown in Fig.~\ref{fig:CHSH}(a). The violation of the classical limit of $S_2$ is clearly observed with a small response time. And the theoretical limit $2\sqrt{2}$ can be achieved with a infinitely small response time. With a larger response time, the measured maximal $S_2$ decreases [Fig.~\ref{fig:CHSH}(a)]. When the response time is larger than the threahold value $T_0$ with $\text{sinc}^2(\delta\omega T_0/2)= 1/\sqrt{2}$ corresponding to $T_0\approx0.32\times(2\pi/\delta\omega)$, the maximal measurable value of $S_2$ falls below the classical limit, meaning the quantum correlation cannot be distinguished.
Here, we assume unity detection efficiency. Lower efficiency will inevitably limits the overall projection measurement efficiency and decrease signal-to-noise ratio in experiments. For certain applications, such as loop-hole free CHSH test~\cite{giustina2013bell,christensen2013detection}, overal efficiency above certain threshold will be required.

{\em{Discussion and Outlook.}}---In addition to eliminating nonlinear optical processes, another major advantage of our approach is scalability. From a theoretical perspective, the systematic approach to design photonic circuit for a quantum frequency comb with an arbitrary large size has been carried out as shown in Fig.~\ref{fig:SU}. From the experimental perspective, the complete photonic circuit can be robustly fabricated on nanophotonic platforms such as silicon and silicon nitride at large scale with standard CMOS technology~\cite{politi2009integrated}. Such a photonic circuit can also be conveniently tuned through thermal-optic effect to realize different projection bases ~\cite{politi2009integrated}. Moreover, superconducting single photon detectors have been integrated on various nanophotonic circuits with high efficiency and low jitter~\cite{pernice2012high,schuck2016quantum}. Therefore, our approach can be scalable fabricated with complete integrated quantum photonic platforms. Extra caution should be taken to minimize the frequency dependence of unitary $U$. One possible approach is to design phase shifters using waveguides with length $L=\frac{2\pi c}{n_g\delta\omega}$ and near-zero dispersion. Furthermore, the achievable length of optical delay line with nanophotonic circuits limits the frequency resolution of the circuits. Currently, nanophotonic optical delay line with length above 30 m and loss below 0.08 dB/m has been demonstrated ~\cite{lee2012ultra}, leading to frequency resolution well below 1 GHz. 

While fidelity of processing frequency-encoded quantum states with our approach depends critically on the response time, the major limiting factor is the detector jitter. Currently, superconducting single photon detectors have achieved a jitter time below 3~ps, with a theoretical limit below 1~ps~\cite{korzh2018demonstrating,allmaras2019intrinsic}. Current jitter performance is enough to maintain fidelity above 90\% for an 85-mode free-space quantum frequency comb, with free spectral range normally below 1 GHz. The maximum frequency range that can preserve quantum effect is determined by the two-mode CHSH test, $\delta\omega/2\pi\approx0.32/T\approx$100 GHz, which is accessible by most integrated photonic platforms.

Finally, a few remarks are in order on potential uses of our proposed method for continuous-variable (CV) photonic quantum information processing. While our derivation is based on single-photon state, the extension to arbitrary multiphoton input states with photon-number-resolving (PNR) detectors is straightforward. This is of critical importance for CV quantum information processing. One possible application with high impact will be Gaussian-Boson-Sampling (GBS)~\cite{hamilton2017gaussian}. We start with N-mode CV Gaussian quantum frequency comb generated with an optical parametric oscillator (OPO), and use our proposed method to implement projection measurement with fast PNR detection on all output ports. In this case, GBS can be realized in frequency domain with very compact device and high efficiency~\cite{pfister2019continuous,menicucci2006universal,weedbrook2012gaussian}.

In conclusion, we have proposed a novel approach to realize high-dimensional quantum information processing encoded in photon frequency domain with passive photonic circuit and time-resolving detection. This approach features no active nonlinear optical process, high scalability, and no intrinsic loss. The capability to process high-dimensional frequency-encoded quantum photonic states can be dramatically improved, benefiting critical quantum applications including cluster-model quantum computing~\cite{zhou2003quantum}, high-dimensional quantum correlation verification~\cite{kaszlikowski2000violations,highdimensionCHSH}, and high-efficiency Boson sampling~\cite{laibacher2018toward}.

\paragraph*{Acknowledgments}This work was supported by the National Science Foundation Grant No. ECCS-1842559 and No. CCF-1907918. C. C thanks Dr. Zheshen Zhang for helpful discussion.

\bibliography{Ref}

\begin{thebibliography}{46}
\bibitem[{Com()}]{CombCondition}
\bibinfo{note}{The design shown in Fig~\ref{fig:SU}b assumes that the center
  frequency $\omega_0$ is a multiple of the free-spectral-range $\delta\omega$,
  which is true for an optical cavity with zero dispersion. For a more general
  case where $\omega_0$ is not a multiple of $\delta\omega$, we could set a
  minimum frequency resolution $\gamma$, and a natural choice would be the
  linewidth of frequency modes. Therefore, we can always find a pair of coprime
  integer $P$ and $Q$ and two frequency detunings $\Delta_0$ and $\Delta$, that
  satisfy $Q(\omega_0+\Delta_0)=P(\delta\omega+\Delta)$ with $\Delta_0<\gamma$
  and $\Delta<\gamma$. Therefore, the time delay for MZIs at $j$th level is
  changed to $\tau_j=Q\pi/(2^{(j-1)}\delta\omega)$, which can separate
  different frequency modes into different paths within the resolution.}
\end{thebibliography}

\end{document}